\documentclass[twocolumn,showpacs,preprintnumbers,
amsmath,amssymb]{revtex4-1}
\usepackage{graphicx,hyperref}
\usepackage{multirow}

\def\beq{\begin{equation}}
\def\eeq{\end{equation}}
\def\bea{\begin{eqnarray}}
\def\eea{\end{eqnarray}}
\def\ba{\begin{array}}
\def\ea{\end{array}}
\def\bit{\begin{itemize}}
\def\eit{\end{itemize}}

\def\nn{\nonumber}

\def\si{\sigma}

\def\cC{{\cal C}}

\newcommand{\cN}{\mathcal{N}}

\def\GB{{{\rm E}_4}}

\newcommand{\rf}[1]{(\ref{#1})}

\makeatletter
\def\underbracket{%
  \@ifnextchar [ %
    {\@underbracket}%
    {\@underbracket [\@bracketheight]}}
\def\@underbracket[#1]{%
  \@ifnextchar [ %
    {\@under@bracket[#1]}%
    {\@under@bracket[#1][0.4em]}}
\def\@under@bracket[#1][#2]#3{
  \mathop {%
    \vtop {%
      \m@th \ialign {%
        ##\crcr $\hfil \displaystyle {#3}\hfil $%
       \crcr \noalign %
       {\kern 3\p@ \nointerlineskip }%
        \upbracketfill {#1}{#2}
       \crcr \noalign %
       {\kern 3\p@ }%
     }%
   }%
  }%
  \limits%
}
\def\upbracketfill#1#2{%
  $\m@th \setbox \z@ \hbox {$\braceld$}
  \edef\@bracketheight{\the\ht\z@}\bracketend{#1}{#2}
  \leaders \vrule \@height #1 \@depth \z@ \hfill
  \leaders \vrule \@height #1 \@depth \z@ \hfill%
  \bracketend{#1}{#2}$%
}
\def\bracketend#1#2{\vrule height #2 width #1\relax}
\begin{document}
\title{Cancellation of   Conformal and Chiral  Anomalies in $\cN\geq 5$ supergravities}
\author{Renata Kallosh }
\affiliation{SITP and Department of Physics, Stanford University, Stanford, CA
94305, USA
}

\begin{abstract} 

A  cancellation of conformal anomalies in $d=4$ for  the 
$C^2+\bar C^2$  and of chiral anomalies for the $C^2-\bar C^2$ is known in  $\cN$-extended supergravities with $\cN \geqslant 5$.
We propose an explanation of these cancellations using supersymmetry and dimension of linearized chiral superfields. We  contrast these models with the ones with $\cN < 5$, where both types of anomalies are known to be present.
\end{abstract}

\maketitle

\vspace{0.1cm}
\section{Introduction} 
 
We study anomalies in d=4 $\cN$-extended supergravities. Anomalies are relevant for the understanding of the UV properties in these theories. This relation was demonstrated for  $\cN=4$ supergravity where the finite $U(1)$ anomalous 1-loop amplitudes were found in \cite{Carrasco:2013ypa}. These same structures have showed up as 4-loop UV divergences in the computations in \cite{Bern:2013uka}.
For $\cN\geq 5$ supergravities the first step in this direction is to explain the known conformal  and chiral  anomaly cancellation  \cite{Nicolai:1980td,Marcus:1985yy,Meissner:2016onk}  even before the structure of anomalous amplitudes is known.
 The underlying studies of conformal anomalies were performed in \cite{Deser:1976yx,Duff:1977ay,Christensen:1978gi,Christensen:1978md,Eguchi:1980jx,Duff:1982yw,Fradkin:1981jc,Fradkin:1982bd,Fradkin:1985am,Deser:1993yx,Duff:1993wm,Komargodski:2011vj,Tseytlin:2013jya}.  In theories in which classical theory does not have Weyl symmetry, like Poincar\'e supergravity, the term `conformal anomaly' is still used as defined in \cite{Duff:1993wm}; it describes  the difference between $g^{\mu\nu} \langle T_{\mu\nu}\rangle_{\rm reg}$ and $\langle g^{\mu\nu} T_{\mu\nu}\rangle_{\rm reg}$. The last term vanishes when classical Weyl symmetry is present.

\vspace{0.1cm}

\noindent{\bf 1. Conformal  Anomaly Status.}  In four dimensions,  the anomaly is often used in the form proposed in
\cite{Duff:1993wm}, 
\beq
{\cal A}_{\rm conf}=T^\mu_{\; \; \mu}=\frac{1}{180(4\pi)^2}\left(c_{s}\cC^2+a_s \GB \right) ,
\label{anom}
\eeq\\[-4mm]
where 
\bea\label{CGB}
\cC^2&\equiv&\cC_{\mu\nu\rho\si}\cC^{\mu\nu\rho\si}=R_{\mu\nu\rho\si}
R^{\mu\nu\rho\si}-2R_{\mu\nu}R^{\mu\nu}+\frac13 R^2 \ ,\nn\\
\GB &=& R^* R^*= R_{\mu\nu\rho\si}
R^{\mu\nu\rho\si}-4R_{\mu\nu}R^{\mu\nu}+R^2 \ ,
\eea
and the coefficients $c_s$ and $a_s$ depend on the spin $s$ of the fields that
couple to gravity.
Here $\GB$ is the Gauss-Bonnet density, a total derivative that gives a topological invariant, Euler number,
when integrated over a 4-dimensional manifold. 
$\cC^2$ is the square of the Weyl tensor. One can also use the following relation 
$
C^2= R^* R^* +2(R_{\mu\nu}^2- {1\over 3} R^2)$.
It shows that in the Einstein space with $R_{\mu\nu}=0$, which corresponds to on-shell gauge-independent supergravity, there is only one contribution to anomaly, $(c_s+ a_s) R_{\mu\nu\rho\si}
R^{\mu\nu\rho\si}$. However, the values of the coefficients $a_s$ and $c_s$  were computed separately in \cite{Christensen:1978md}.
For the gauge fields, vectors, gravitini and graviton, the computations were performed in the harmonic gauge: $\nabla^\mu A_\mu= \gamma^\mu \psi^\mu= \nabla^\mu (h_{\mu\nu} - {1\over 2} g_{\mu\nu} h^\rho{}_\rho)=0$, and the results are known  in a general gravitational background, as given in the Table 1. below. 

In the supersymmetry context it is convenient to use the two-component  Weyl spinors,
\beq
\cC_{\mu\nu\rho\si}\cC^{\mu\nu\rho\si} \Rightarrow  C_{\alpha \beta \gamma\delta} C^{\alpha \beta \gamma\delta} +\bar C_{\dot \alpha \dot \beta \dot \gamma \dot \delta} \bar C^{\dot \alpha \dot \beta \dot \gamma \dot \delta}\equiv C^2+\bar C^2\nonumber
\eeq

\vspace{1mm}

Following \cite{Meissner:2016onk}, we survey supergravities
for which the contribution from the different spins to  the coefficients $c_s$ and 
$a_s$ cancel.
Table 1 shows the coefficients  $c_s$ and $a_s$ coming from the integration 
over massless  fields  with spins 
up to $s\!=\!2$. 
The entry labeled $0^*$ give the result
for two-form field; it gives the same contribution 
to $c_0$ as the scalars, but its contribution to the $a_0$ coefficient is different.

  As shown in  \cite{Nicolai:1980td}, 
the sum $\sum (c_s + a_s)$  vanishes for all $\cN = 3$ supermultiplets with
maximum spin $s\leqslant 2$ when
 one scalar is replaced by one two-form field in the spin-$\frac32$ multiplet; the resulting
multiplets can then be used as building blocks to arrange for all $N\geqslant 4$ supergravities 
to have vanishing $\sum (c_s + a_s)$. 

\renewcommand{\arraystretch}{1.3}
\begin{center}
\begin{tabular}{|c||c|c||}
\hline
\multirow{2}{*} 
& $c_s$ & $a_s$\\ \hline \hline
\  $0$($0^*$) &\ $\frac32$$(\frac32)\ $ &\ $-\frac12$($\frac{179}2$)  \\[3pt] \hline
$\frac12$ & $\frac92$  & $-\frac{11}{4}$  \\[1pt] \hline
$1$ & $18$  & $-31$   \\[1pt] \hline
$\frac32$ & $-\frac{411}{2}$   & $\frac{589}{4}$    \\[1pt] \hline
$2$ & $783$   & $-571$     \\[1pt] \hline
\end{tabular}
\vspace{2mm}

Table 1. Coefficients of the conformal anomaly in \cite{Meissner:2016onk}. 
\end{center}

The separate $\cC^2$ anomaly was studied  recently for extended supergravities in \cite{Meissner:2016onk}. It was shown that 
 $\cN\leqslant 4$ supergravities cannot have  cancellation, neither in pure supergravities, nor in models  with matter. For example, in pure $\cN=4$ supergravity
$
{\cal A}_{\rm conf}^{\cN=4}=  c_2 + 4 c_{\frac32} + 6 c_1 + 4 c_{\frac12} + 2 c_0 = 90$.  However, it was  found that the total contribution $\sum c_s$ vanishes for  $\cN \geqslant 5$ supergravities. Only gravitinos give a negative contribution. Therefore starting with $\cN=5$ the cancellation is possible and actually takes place:
\bea\label{5}
{\cal A}_{\rm conf}^{\cN=5}=  c_2 + 5 c_{\frac32} + 10 c_1 + 11 c_{\frac12} + 10 c_0 &=& 0\; ,\\
\label{6}
{\cal A}_{\rm conf}^{\cN=6}= c_2 + 6 c_{\frac32} + 16 c_1 + 26 c_{\frac12} + 30 c_0 &=& 0 \; ,\\
\label{8}
{\cal A}_{\rm conf}^{\cN=8}= c_2 + 8 c_{\frac32} + 28 c_1 + 56 c_{\frac12} + 70 c_0 &=& 0 \;.
\eea
None of these models have matter multiplets, only pure supergravity is available for $\cN=5,6,8$, $\cN=7$ is equivalent to $\cN=8$.

\vspace{3mm}

\noindent{\bf 2. Chiral Anomaly Status.} 
$\cN$-extended supergravities of interest have scalars in the coset space ${G\over H}$, where the isotropy group $H$ is $U(4), U(5), U(6), SU(8)$, for $\cN=4,5,6,8$,  respectively. The models where $H$ has a $U(1)$ subgroup, potentially have a $U(1)$ chiral anomaly, which was computed in \cite{Marcus:1985yy}. 
The gravitational part of $U(1)$ chiral anomaly is proportional to $R_{\mu\nu\rho\si}\tilde  R^{\mu\nu \rho \si}$ and can be given in the form
\bea
{\cal A}_{\rm chiral}&=& \partial^\mu J_{\mu}^5=b_{h}(C_{\alpha \beta \gamma\delta} C^{\alpha \beta \gamma\delta} -\bar C_{\dot \alpha \dot \beta \dot \gamma \dot \delta} \bar C^{\dot \alpha \dot \beta \dot \gamma \dot \delta}) \equiv\cr
\cr
&& \equiv b_h  (C^2-\bar C^2) \ ,
\label{anom2}
\eea\\[-4mm]
where the coefficients $b_h$  depend on helicity.
It was found in \cite{Marcus:1985yy} that the $U(1)$ chiral anomaly contribution to $\cN=5,6$ supergravity cancels between the members of the supermultiplet.

In $\cN=5$ there are  5 gravitini $\psi^i_\mu$, 10 chiral vectors $F^{ij}$,  10 spin 1/2 fields $\chi^{ijk}$,  a singlet  spin 1/2 field $\chi$, and 10 scalars. In  $\cN=6$ there are 6 gravitinos $\psi^i_\mu$, 15 chiral vectors $F^{ij}$, and a singlet chiral vector $F$, 20 spin 1/2 fields $\chi^{ijk}$, and 6 spin 1/2 fields $\chi^i$, and 30 scalars. Therefore one finds, using Table 3 in \cite{Marcus:1985yy}, that
\beq \label{5chiral}
 {\cal A}_{\rm chiral}^{\cN=5}=  -21\times 5 + 4\times 10\times 2  + 10\times 3 -5=0\ , 
 \eeq  
\bea \label{6chiral}
{\cal A}_{\rm chiral}^{\cN=6}=&-& 21\times 6 + 4\times 15\times 2 + 4\times (-6 )+\nonumber  \\
\cr
& + & 20\times 3 + 6\times (-5)=0 \ .
\eea
 $\cN=8$ supergravity does not have a chiral $U(1)$ anomaly since $SU(8)$ does not have an $U(1)$ subgroup.
Other  types of triangle anomalies in $\cN=4,5,6,8$ supergravities are associated with the $SU(\cN)$ subgroups and were also computed in \cite{Marcus:1985yy}. The relevant  currents $J^{\mu i}$ for $\cN\geq 5$ were also found to be anomaly-free. For example,  the $SU(8)$ anomaly in $\cN=8$ supergravity is also cancelled when vectors and spinors are taken into account.

\vspace{3mm}

\section{Supersymmetry}
We would like to explain the fact that in $\cN=5,6,8$ supergravities both conformal as well as chiral anomalies vanish.
Both  anomalies can be described using superfields, following  \cite{Ferrara:1977mv,Kallosh:1980fi}. We use the linearized superfields below. We have seen in dealing with 1-loop anomalies in $\cN=4$ supergravity in \cite{Carrasco:2013ypa} using both superfields and superamplitudes, that using linearized superfields was efficient and predictive. The consequent computation of UV divergences in four loops in \cite{Bern:2013uka} has revealed the UV divergences which exist only in terms of superspace integrals for linearized superfields. 
Note that the superamplitudes have manifest linearized supersymmetry of asymptotic on-shell physical states, which might explain why linearized superfields are useful.

\vspace{3mm}

1. {\bf In  $\cN=1$ supergravity} one finds that  $T^\mu_{\; \; \mu}$ and $ \partial^\mu J_{\mu}^5$ can be given in the superfield form 
\beq\label{N1}
 T^\mu_{\; \; \mu} \Rightarrow \int d^2 \theta \, W_{\alpha\beta \gamma} W^{\alpha\beta \gamma} +h.c. = C_{\alpha \beta \gamma\delta} C^{\alpha \beta \gamma\delta} +h.c.
\eeq

\beq
\partial^\mu J_{\mu}^5 \Rightarrow \int d^2 \theta \, W_{\alpha\beta \gamma} W^{\alpha\beta \gamma} -h.c. = C_{\alpha \beta \gamma\delta} C^{\alpha \beta \gamma\delta} -h.c.
\eeq

2. {\bf In  $\cN=2$ supergravity} one finds 
\beq
 T^\mu_{\; \; \mu} \Rightarrow \int d^4 \theta \, W_{\alpha\beta } W^{\alpha\beta } +h.c. = C_{\alpha \beta \gamma\delta} C^{\alpha \beta \gamma\delta} +h.c.
\eeq

\beq
\partial^\mu J_{\mu}^5 \Rightarrow \int d^4 \theta \, W_{\alpha\beta } W^{\alpha\beta } -h.c. = C_{\alpha \beta \gamma\delta} C^{\alpha \beta \gamma\delta} -h.c.
\eeq

3. {\bf In  $\cN=3$ supergravity} 
\beq
 T^\mu_{\; \; \mu} \Rightarrow \int d^4 \theta \, W_{\alpha } W^{\alpha } +h.c. = C_{\alpha \beta \gamma\delta} C^{\alpha \beta \gamma\delta} +h.c.
\eeq

\beq
\partial^\mu J_{\mu}^5 \Rightarrow \int d^4 \theta \, W_{\alpha } W^{\alpha } -h.c. = C_{\alpha \beta \gamma\delta} C^{\alpha \beta \gamma\delta} -h.c.
\eeq

4. {\bf In  $\cN=4$ supergravity}  
\beq
 T^\mu_{\; \; \mu} \Rightarrow \int d^8 \theta \, W ^2+h.c. = C_{\alpha \beta \gamma\delta} C^{\alpha \beta \gamma\delta} +h.c.
\eeq

\beq \label{N4}
\partial^\mu J_{\mu}^5 \Rightarrow \int d^8 \theta \, W^2 -h.c. = C_{\alpha \beta \gamma\delta} C^{\alpha \beta \gamma\delta} -h.c.
\eeq

Thus, for $\cN\leq 4$ there are candidates for conformal and chiral anomalies with correct dimension (4 in each case) which agree with supersymmetry.

By the time we reach $\cN \geq 5$, it seems that we have run out of luck, as  we do not expect linearized chiral superfields there. Surprisingly, 
this is actually not true, there are linearized chiral superfields in $\cN\geq 5$, which can be used to construct the susy version of conformal and chiral anomalies. The chiral superfields are $H$-singlets. The details on these superfields will be presented  elsewhere.

\subsection{C-superfields for all $\cN$}\label{C}
It was argued in \cite{Carrasco:2013ypa} that in $\cN=4$ supergravity there is a chiral superfield $\bar C_{\dot \alpha \dot \beta \dot \gamma \dot \delta} (x, \theta)$. 
\beq 
D_{\dot \eta i}\,  \bar C_{\dot \alpha \dot \beta \dot \gamma \dot \delta} (x, \theta, \bar \theta)=0\, ,  \quad D_{ \eta}^{ i}\,   C_{ \alpha  \beta \gamma  \delta} (x, \theta,\bar \theta)=0 \ .
\label{Csuperfield}\eeq
The same argument as in \cite{Carrasco:2013ypa} works for $\cN \geq 5$. One can therefore look for the candidates for conformal-chiral 1-loop anomaly for all $\cN$ in the form
\beq
\int d^{2\cN} \theta \, \bar C_{\dot \alpha \dot \beta \dot \gamma \dot \delta} \, \bar C^{\dot \alpha \dot \beta \dot \gamma \dot \delta}  \pm h.c.
\eeq
However,  the dimension of this expression is equal to $\cN+4$, whereas the 1-loop candidates for  $T^\mu_{\; \; \mu}$ and $\partial^\mu J_{\mu}^5$ have dimension 4. Therefore these are not acceptable for any $\cN \geq 1$.

\subsection{Chiral  superfields in $\cN= 5$}
The set of spin 1/2 fields in $\cN= 5$ model  \cite{deWit:1981yv,Howe:1981gz} includes 10 $\chi^{ijk}$ in the antisymmetric representation of $U(5)$ and a singlet $\chi =\chi^{ijklm}$. It originates from $\cN=8$ as a component $\chi^{678}$. In the linear approximation one finds that the singlet spinor is an anti-chiral superfield, and its conjugate is a chiral superfield.
\beq
D_{\alpha}^i \chi_{\beta} =0\, ,  \qquad \bar D_{\dot \alpha i} \bar \chi_{\dot \beta} =0 \ ,
\eeq
so that
\bea
\bar  \chi_{\dot \alpha} (x, \theta)= \bar  \chi_{\dot \alpha} (x) + \theta_i^\alpha \partial_{\dot \alpha \alpha} \phi^i (x) + 
\cdots \cr
\cr
+ \theta_i^\alpha \theta_j^\beta \theta_k^\gamma  \theta_l^\delta \theta_m^\sigma \epsilon^{ijklm} \partial_{\dot \alpha \alpha} C_{\beta\gamma\delta\sigma} \ .
\eea
Now we are in a position to present a superspace analog of bosonic currents $T^\mu_{\; \; \mu}$ and $\partial^\mu J_{\mu}^5$, as we did for smaller $\cN$.
\bea
  T^\mu_{\; \; \mu} \quad &\Rightarrow& \quad \int d^{10} \theta \, \bar  \chi_{\dot \alpha}^2 (x, \theta) +h.c. \cr
\cr
 &=&\partial_{\dot \alpha \sigma} \, C_{\alpha \beta \gamma\delta}  \, \partial^{\dot \alpha \sigma}  \, C^{\alpha \beta \gamma\delta} +h.c.
\eea
\bea
\partial^\mu J_{\mu}^5 \quad &\Rightarrow& \quad \int d^{10} \theta \, \bar  \chi_{\dot \alpha}^2 (x, \theta) -h.c. \cr
\cr
&=& \partial_{\dot \alpha \sigma} \, C_{\alpha \beta \gamma\delta}  \, \partial^{\dot \alpha \sigma}  \, C^{\alpha \beta \gamma\delta} -h.c.
\eea
These are not supporting 1-loop anomalies, as they have dimension 6. The absence of the relevant supersymmetric version of conformal/chiral anomalies $(C_{\alpha \beta \gamma\delta}  \,  C^{\alpha \beta \gamma\delta} \pm h.c.)$ explains the cancellation of both conformal and chiral 1-loop anomalies, as shown in eqs. \rf{5} and \rf{5chiral}.
There are no other chiral superfields and integrals over chiral subspaces which would support 1-loop anomalies in agreement with linearized supersymmetry.

\subsection{Chiral  superfields in $\cN= 6$}
The set of spin 1 fields in $\cN= 6$ model  \cite{Howe:1981gz} includes 15 of them in
 antisymmetric representation of $U(6)$, $F^{ij}$,  and an $U(6)$ singlet $F= F^{ijklmn}$, which originates from $\cN=8$ as a component $F^{78}_{\alpha \beta}, \bar F_{78}^{\dot \alpha \dot \beta}$. One finds
that
\beq
D_{\dot \alpha i} \bar M_{\dot \beta \dot \gamma} =0\ , \qquad  D_{ \alpha}^{ i} M_{ \beta  \gamma} =0 \ ,
\eeq
\bea
\bar M_{\dot \alpha \dot \beta} (x, \theta)= \bar M_{\dot \alpha \dot \beta}(x) + \theta_i^\alpha \partial_{\dot \alpha \alpha} \chi^i _{ \dot \beta}(x) + 
\cdots \cr
\cr
+ \theta_i^\alpha \theta_j^\beta \theta_k^\gamma  \theta_l^\delta \theta_m^\sigma  \theta_n^\eta \epsilon^{ijklmn} \partial_{\dot \alpha \alpha} \partial_{\dot \beta \beta} C_{\gamma\delta\sigma\eta} \ .
\eea
Now again we are in a position to present a superspace analog of bosonic currents $T^\mu_{\; \; \mu}$ and $\partial^\mu J_{\mu}^5$, as we did for smaller $\cN$.
\bea
 T^\mu_{\; \; \mu} \quad &\Rightarrow& \quad \int d^{12} \theta \, \bar  M_{\dot \alpha \dot \beta }^2 (x, \theta) +h.c. \cr
\cr
&=&   \partial_{\dot \delta \zeta} \partial_{\dot \alpha \sigma} \, C_{\alpha \beta \gamma\delta}  \, \partial^{\dot \alpha \sigma}   \, \partial^{\dot \delta \zeta}\, C^{\alpha \beta \gamma\delta} +h.c.
\eea
\bea
 \partial^\mu J_{\mu}^5 \quad &\Rightarrow& \quad \int d^{12} \theta \, \bar  M_{\dot \alpha \dot \beta }^2 (x, \theta) -h.c. \cr
\cr
&=&   \partial_{\dot \delta \zeta} \partial_{\dot \alpha \sigma} \, C_{\alpha \beta \gamma\delta}  \, \partial^{\dot \alpha \sigma}   \, \partial^{\dot \delta \zeta}\, C^{\alpha \beta \gamma\delta} -h.c.
\eea
As in $\cN=5$, these are not supporting 1-loop anomalies, they have dimension 8. The absence of the relevant supersymmetric version of conformal/chiral anomalies explains the cancellation of both conformal and chiral 1-loop anomalies, as shown in eqs. \rf{6} and \rf{6chiral}.

\subsection{Chiral  superfields in $\cN= 8$}
As suggested in sec. IIA, there is a linearized chiral C-superfield \rf{Csuperfield}, which has the form
\bea
&& \bar C_{\dot \alpha \dot \beta \dot \gamma \dot \delta} (x, \theta)=  \bar C_{\dot \alpha \dot \beta \dot \gamma \dot \delta} (x)+\cdots +\cr
\cr
 &+&\theta^{\alpha_1}_{i_1} \theta^{\alpha_2}_{i_2} \theta^{\alpha_3}_{i_3} \theta^{\alpha_4}_{i_4} \theta^{\alpha}_{j_1} \theta^{\beta}_{j_2} \theta^{\gamma}_{j_3} \theta^{\delta_1}_{j_4}\epsilon ^{i_1 i_2 i_3 i_4 j_1 j_2 j_3 j_4} \cr
 \cr
&\times& \partial_{\dot \alpha \alpha_1}  \partial_{\dot \beta \beta_1}  \partial_{\dot \gamma  \gamma_1}  \partial_{\dot \delta \delta_1}   \, C_{\alpha\beta \gamma\delta}
\eea
The relevant integral is
\bea
&&\int d^{16} \theta \, \bar C_{\dot \alpha \dot \beta \dot \gamma \dot \delta} (x, \theta) \bar C^{\dot \alpha \dot \beta \dot \gamma \dot \delta} (x, \theta)= \cr
\cr
&& \partial_{\dot \alpha}^{ \alpha_1}  \partial_{\dot \beta}^{ \beta_1}  \partial_{\dot \gamma}^{ \gamma_1}  \partial_{\dot \delta}^{ \delta_1}   \, C_{\alpha\beta \gamma\delta}  \partial^{\dot \alpha}_{ \alpha_1}  \partial^{\dot \beta}_{ \beta_1}  \partial^{\dot \gamma}_{ \gamma_1}  \partial^{\dot \delta}_{ \delta_1}   \, C^{\alpha\beta \gamma\delta} \ .
\eea
It has indeed dimension $12$, which is $4+\cN$ for $\cN=8$. This explains the cancellation of conformal anomalies in eq. \rf{8} for $\cN=8$ supergravity. There are no other linearized superfields.

\section{Conclusion}

In conclusion, our goal is achieved, we explained why dimension 4 operators describing anomalies are absent in $\cN=5,6,8$ supergravity in d=4:
\beq
{\cal A}_{\rm conf}=T^\mu_{\; \; \mu}= {\cal A}_{\rm chiral}=\partial^\mu J_{\mu}^5=0  \ .
\eeq


 We have explained the zeros in eqs. \rf{5}, \rf{6}, \rf{8}, \rf{5chiral}, \rf{6chiral} as a consequence of supersymmetry/dimension. The absence of 1-loop conformal/chiral anomalies is due to dimension of  linearized chiral superfields, which are present in $\cN=5, 6, 8$  models. Same methods in models with $\cN\leq 4$ are shown to provide candidate anomalies for the 1-loop models, which is in agreement with computations performed in \cite{Marcus:1985yy} and \cite{Meissner:2016onk}. For all $\cN\leq 4$ supergravities, the dimension of relevant chiral superfields is such that the supersymmetric anomaly candidates for $ T^\mu_{\; \; \mu}(x)$ and  $ \partial^\mu J_{\mu}^5$ have dimension 4, as required for the 1-loop anomaly in d=4. In  $\cN \geq  5$ supergravities, these dimensions exceed 4, which is the reason for anomaly cancellation. 
 
 Our analysis supports the earlier  argument in \cite{deWit:1978pd} based on the spin counting in the action which include both the   Poincare as well as  Conformal Supergravity. The fact that we have found here only invariants with the  dimension exceeding 4 for $\cN \geq  5$ supergravities means that Weyl square, which has dimension 4, has no supersymmetric completion. This corresponds to the fact discovered in \cite{deWit:1978pd} that there are no massive multiplets with top
spin 2 for $\cN \geq  5$.

Interesting new questions can be adressed  with regard to conformal anomalies studied here.  What is the role of the harmonic gauge, where eqs.   \rf{5}, \rf{6}, \rf{8} are valid in a general gravitational background? Is it a gauge consistent with supersymmetry?  What is the relation to anomalies in conformal supergravities?

It would be  also interesting to develop the analysis of this paper to study the anomalies in the on-shell amplitudes in $\cN\geq 5$ supergravities, following $\cN=4$ case 
in  \cite{Carrasco:2013ypa,Bern:2013uka}. It is particularly important in view of the fact that $\cN= 5$ supergravity is UV finite at 4 loops as was shown in \cite{Bern:2014sna}.

\vskip 1 mm

\noindent{\bf {Acknowledgments:}}  I am grateful to Z. Bern and H. Nicolai for stimulating discussions, and to  
S. Ferrara,  D. Freedman,   R. Roiban, A. Tseytlin, B. de Wit and  Y. Yamada for their important feedback to this work.  This work  is supported by SITP and by the NSF Grant PHY-1316699.

\end{document}